\documentclass[fleqn,11pt,letterpaper]{article}
\usepackage{amsfonts}
\usepackage{amssymb}
\usepackage{amsmath}
\usepackage{amsthm}
\usepackage{enumerate}
\usepackage{multirow}
\usepackage{float}
\usepackage{graphicx}
\usepackage{epstopdf}
\usepackage{color}
\usepackage[margin=2.57cm]{geometry}
\definecolor{darkred}{rgb}{0.5,0.2,0.2}
\usepackage[pdftitle={Income and wellbeing},pdfauthor={Mikkel Bennedsen},colorlinks=TRUE,allcolors=darkred]{hyperref}
\usepackage{booktabs}
\usepackage{pdflscape}
\usepackage[round]{natbib}

\usepackage{tabularx}

\usepackage{float}
\floatstyle{plaintop}
\restylefloat{table}

\restylefloat{table}

\theoremstyle{plain}

\def\bi{\begin{itemize}}
\def\ei{\end{itemize}}

\allowdisplaybreaks
\graphicspath{{Figures/}}

\newif\ifi

\title{Income and emotional well-being: \\ Evidence for well-being plateauing around \$200,000 per year}

\author{Mikkel Bennedsen\thanks{
Department of Economics and Business Economics, 
Aarhus University, 
Fuglesangs All\'e 4,
8210 Aarhus V, Denmark.
E-mail:\
\href{mailto:mbennedsen@econ.au.dk}{\nolinkurl{mbennedsen@econ.au.dk}} 
}}

\begin{document}

\maketitle

\begin{abstract}
Is emotional well-being monotonically increasing in the level of income or does it  reach a plateau at some income threshold, whereafter additional income does not contribute  to further well-being? Conflicting answers  to this question has been suggested in the academic literature. In a recent paper, using an income threshold of \$100,000 per year,  \cite{KKM2023} appears to have resolved these conflicts, concluding that emotional well-being is monotonically increasing in income for all but the unhappiest individuals. In this paper, we show that this conclusion is sensitive to the placement of the income threshold at which the relationship between emotional well-being and income is allowed to plateau. Using standard econometric methods, we propose a data-driven approach to detect the placement of the threshold. Using this data-driven income threshold, a flat relationship between household income and emotional well-being above a threshold around \$200,000 per year is found. While our analysis relaxes the assumption of a pre-specified income threshold, it relies on a number of other assumptions, which we briefly discuss. We conclude that although the analysis of this paper provides some evidence for well-being plateauing around \$200,000 per year, more research is needed before any definite conclusions about the relationship between emotional well-being and income can be drawn. 

\vspace*{0.5cm}


\end{abstract}

\newpage

\section{Introduction}
Is well-being monotonically increasing in the level of income or does it  reach a plateau at some income threshold? The answer to this question has important implications for both personal choice and public policy.  Two measures of well-being are generally distinguished between in the literature, namely \emph{emotional} well-being and \emph{evaluative} well-being \citep[e.g.][]{Diener2010}. Emotional well-being refers to how an individual experiences their everyday life, while evaluative well-being refers to how an individual evaluates her own life as a whole. It has been found that evaluative well-being is monotonically increasing in the level of income \citep[e.g.][]{KD2010}, but the relationship between emotional well-being and income is more disputed. 

In a seminal paper, \cite{KD2010} analyzed a survey of more than 450,000 US citizens and  found that emotional well-being increases linearly with household log-income up to a certain threshold, determined to be around \$75,000 per year, whereafter additional income does not increase well-being further. A similar conclusion was reached in \cite{Jebb2018}. Later, analyzing data sampled in real time using smartphone notifications, \cite{MK2021} re-examined this question and found that emotional well-being increases monotonically with log-income without reaching a plateau. Recently, in a notable instance of ``adversarial collaboration'', Daniel Kahneman and Matthew Killingsworth teamed up to resolve the apparent conflict between these findings \citep{KKM2023}. Using linear regression, \cite{KKM2023} validates the findings of \cite{MK2021}, i.e. that  emotional well-being increases monotonically with log-income without reaching a plateau. Using quantile regression, \cite{KKM2023} finds that emotional well-being does in fact plateau at the threshold \$100,000 per year, but only for the most unhappy individuals in the sample. \citep[\$100,000 being the inflation-adjusted value of the \$75,000 threshold used in][]{KD2010} For the remaining parts of the sample, however, it is again found that emotional well-being increases monotonically with log-income without reaching a plateau. For the happiest individuals, an ``accelerating'' pattern is even found, where the relationship between emotional well-being and log-income becomes even stronger above the threshold. 

The analyses in the above-mentioned papers rest on two crucial assumptions. The first assumption is that the relationship between well-being and log-income is piecewise linear with one change in both intercept and slope. The second assumption is that the change occurs at the income threshold \$100,000 (or \$75,000) per year. In this paper, we examine the effects on the analysis when we retain the first assumption and relax the second. We use standard econometric methods to detect the placement of a structural break in a linear model \citep[e.g.][]{BP1998,perron2006}. Re-doing the linear regression analysis of  \cite{KKM2023} with a data-driven placement of the threshold, we find that emotional well-being does in fact plateau, but the threshold is found to be between \$175,000 and \$250,000, substantially higher than the \$100,000 used in  \cite{MK2021} and \cite{KKM2023}. Then, using the data-driven threshold, we re-do the quantile regression analysis of  \cite{KKM2023} and find evidence of a plateau in the relationship between emotional well-being and log-income, regardless of the part (quantile) of the well-being distribution considered. Although the analysis presented in this paper relaxes a crucial assumption compared to the previous literature, it still imposes a number of quite stringent assumptions.  We close the paper by briefly discussing these assumptions and argue that more research is needed before any definite conclusions about the relationship between emotional well-being and income can be drawn.

\section{Data}
In this paper, we use the data made available by \cite{KKM2023}.\footnote{Data can be found at \url{https://osf.io/qye4a/}, last accessed November 13, 2023. See also \url{https://go.trackyourhappiness.org/} for information about the smartphone app used to collect the data.} These data are constructed from 1,725,944 experience-sampling reports from 33,391 employed adults aged 18 to 65 living in the United States, collected using smartphone notifications to obtain real-time evaluations of experienced (i.e. emotional) well-being. Well-being was reported on a continuous scale between ``Very bad'' and ``Very good'', converted into a number between 0 and 100, and subsequently an average number was constructed for each individual, leading to 33,391 data points. The only explanatory variable used in \cite{KKM2023} was annual pre-tax household income, which was  bracketed into 15 income groups labeled \$15000, \$25000, \$35000, \$45000, \$55000, \$65000, \$75000, \$85000, \$95000, \$112500, \$137500, \$175000, \$250000, \$400000, and \$625000. We refer to \cite{MK2021} for further details on the data.

\section{Empirical analysis}
In Section \ref{sec:OLS} we use ordinary least squares regression to examine the relationship between the conditional mean of emotional well-being   and  log-income. In Section \ref{sec:QR} we use quantile regression to examine the relationship between the conditional quantiles of emotional well-being and log-income. We reproduce the analyses of  \cite{KKM2023}  by inserting a structural break in the linear relationship at the pre-specified threshold \$100,000 per year, and compare with an alternative analysis using a threshold that has been chosen in a data-driven way using standard econometric methods.

\subsection{Ordinary least squares regression analysis}\label{sec:OLS}
We consider the following model
\begin{align}\label{eq:OLS}
E(z_i|x_i) = (a + b x_i) I(\exp(x_i) \leq \tau) + (c+dx_i) I(\exp(x_i) > \tau), \quad i = 1,2,\ldots, 33391,
\end{align}
where $a,b,c,d \in \mathbb{R}$, $z_i$ is a measure of well-being of individual $i$, $x_i$ is the log-income of individual $i$, $E(z_i|x_i)$ is the mean of $z_i$, conditional on $x_i$, and $I(A)$ is the indicator function of the event $A$, i.e. it is equal to one if $A$ is true and zero otherwise. We follow \cite{MK2021} and \cite{KKM2023} and let $z_i$ be the z-scored well-being level reported by individual $i$, i.e. 
\begin{align*}
z_i := \frac{y_i - \overline y}{s_y},
\end{align*}
where $\overline y := N^{-1} \sum_{i=1}^N y_i$ is the sample mean and $s_y^2 := (N-1)^{-1} \sum_{i=1}^N (y_i-\overline y)^2$ is the sample variance of the reported well-being levels, with $N =$ 33,391 being the sample size.

The model for the conditional mean of $z_i$ in \eqref{eq:OLS} is piecewise linear in log-income $x_i$. It features a structural break in both intercept and slope at the threshold value $\tau$; for $\exp(x_i) \leq \tau$ the intercept is $a$ and slope is $b$, while for $\exp(x_i) > \tau$ the intercept is $c$ and slope $d$. This model has been extensively used in the literature \citep[e.g.][]{KD2010,MK2021,KKM2023} but the value for the threshold $\tau$ has arguably been chosen rather arbitrarily. For instance, $\tau$ is set equal to \$100,000 in \cite{MK2021} and \cite{KKM2023}, which is simply the inflation-adjusted value used on a different data set in \cite{KD2010}, where $\tau =$ \$75,000. 

\begin{figure}[t!]
    \centering
        \caption{Ordinary least squares regression results from \eqref{eq:OLS}. Left panel: Sum of squared residuals as a function of the income threshold $\tau$. Middle panel: Threshold at \$100,000. Right panel: Threshold at \$200,000. Standard errors in parentheses. Note: black dots in the middle and right panels are sample averages of well-being levels in each income brackets.}
    \includegraphics[width=0.3\textwidth]{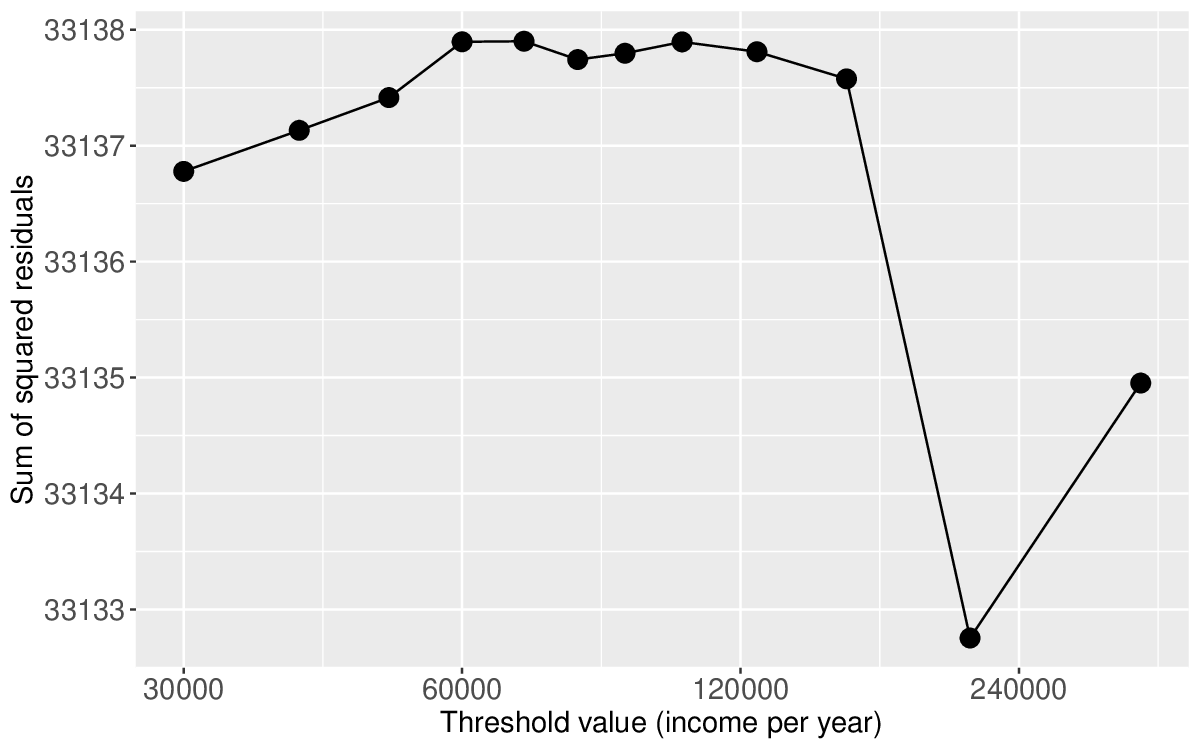}
    \includegraphics[width=0.3\textwidth]{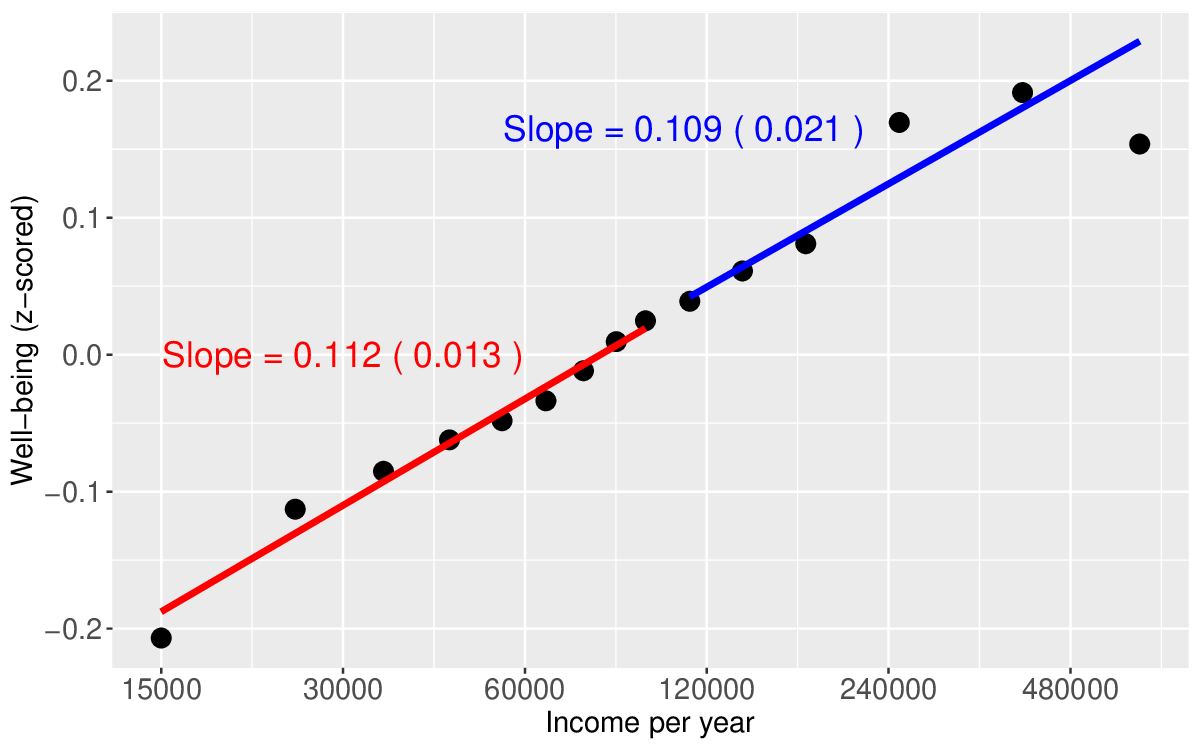}
        \includegraphics[width=0.3\textwidth]{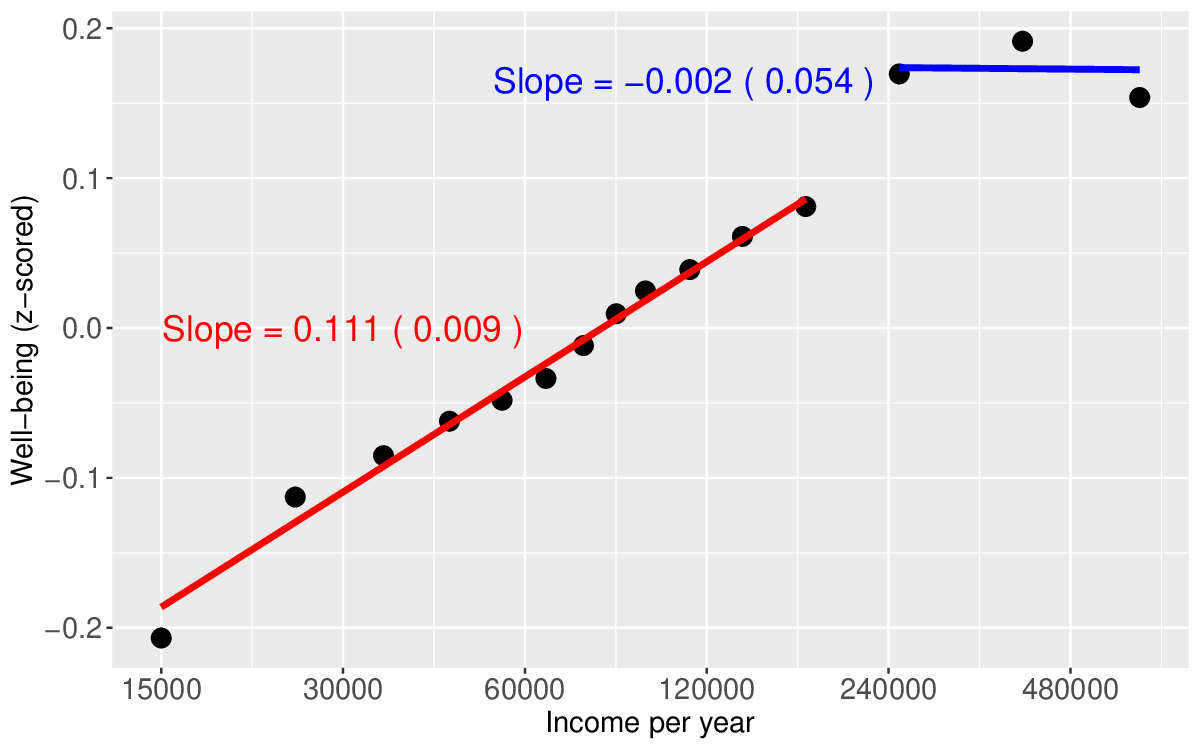}
    \label{fig:full}
\end{figure}

For a given value of $\tau$, the parameters $a,b,c,d$ in \eqref{eq:OLS}  may be estimated via ordinary least squares regression. We follow the econometrics literature on detecting ``structural breaks'' in regression models and propose to choose the threshold $\tau$ as the place where the sum of squared residuals of the regression analysis is minimized \citep{BP1998,perron2006}. The sum of squared residuals for various placements of the threshold are shown in the left panel of Figure \ref{fig:full}. The optimal threshold is found to be between \$175,000 and \$250,000. In the right panel of Figure \ref{fig:full}, we plot the estimated relationship \eqref{eq:OLS} between well-being and log-income using this optimal threshold. For comparison, the middle panel of  Figure \ref{fig:full} presents the estimated relationship using the \$100,000 threshold. The latter is a replication of the results from  \cite{KKM2023} and shows a consistently increasing pattern, but the former shows that choosing the threshold in a data-driven way leads to a flat relationship between income and well-being above the threshold.

\subsection{Quantile regression analysis}\label{sec:QR}
We now consider the following model
\begin{align}\label{eq:QR}
Q_{y_i|x_i}(p) =  (a + b x_i) I(\exp(x_i) \leq \tau) + (c+dx_i) I(\exp(x_i) > \tau), \quad i = 1,2,\ldots, 33391,
\end{align}
where  $a,b,c,d \in \mathbb{R}$, $I(\cdot)$ and $\tau$ as above, and $Q_{y_i|x_i}(p)$ is the $p$'th quantile of $y_i$, conditional on $x_i$, with $p \in (0,1)$. Note that we here follow \cite{KKM2023} and consider the level of well-being $y_i$ as the dependent variable, and not the z-scored version $z_i$ as we did in \eqref{eq:OLS}. Otherwise, the model \eqref{eq:QR} is similar to \eqref{eq:OLS}, except that it is a model of the conditional quantile, instead of the conditional mean, of the well-being measure. 

\begin{figure}[t!]
    \centering
      \caption{Quantile regression results from \eqref{eq:QR}, implemented using the \texttt{quantreg} package in R \citep{Koenker2019}. Left panel: Threshold at \$100,000. Right panel: Threshold at \$200,000. $t$-stats in parentheses, calculated using the quantile regression sandwich formula and the Hall-Sheather
bandwidth rule \citep[e.g. Section 3.2.3 in][]{Koenker2005}. Note: black dots are sample quantiles of well-being levels in each income brackets.}
    \includegraphics[width=0.475\textwidth]{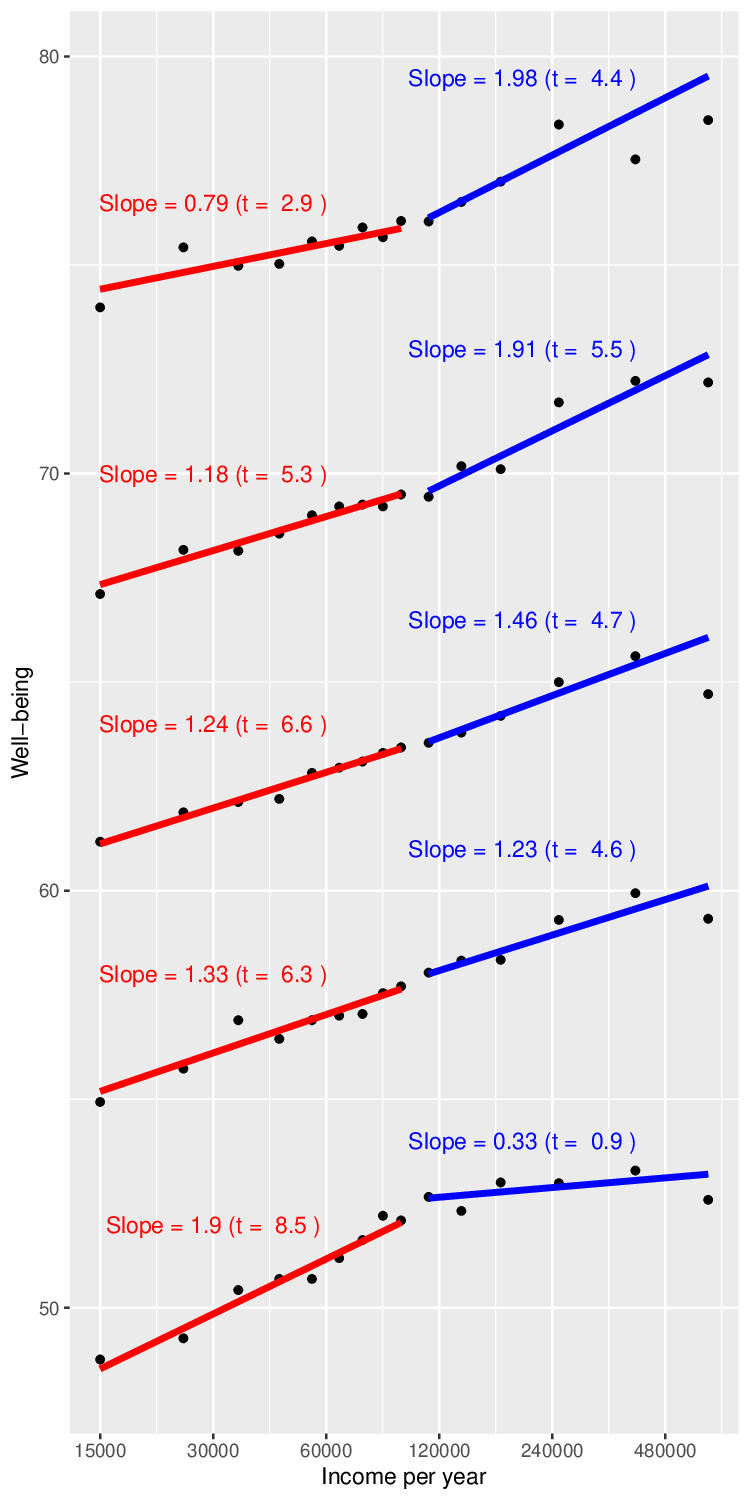}
        \includegraphics[width=0.475\textwidth]{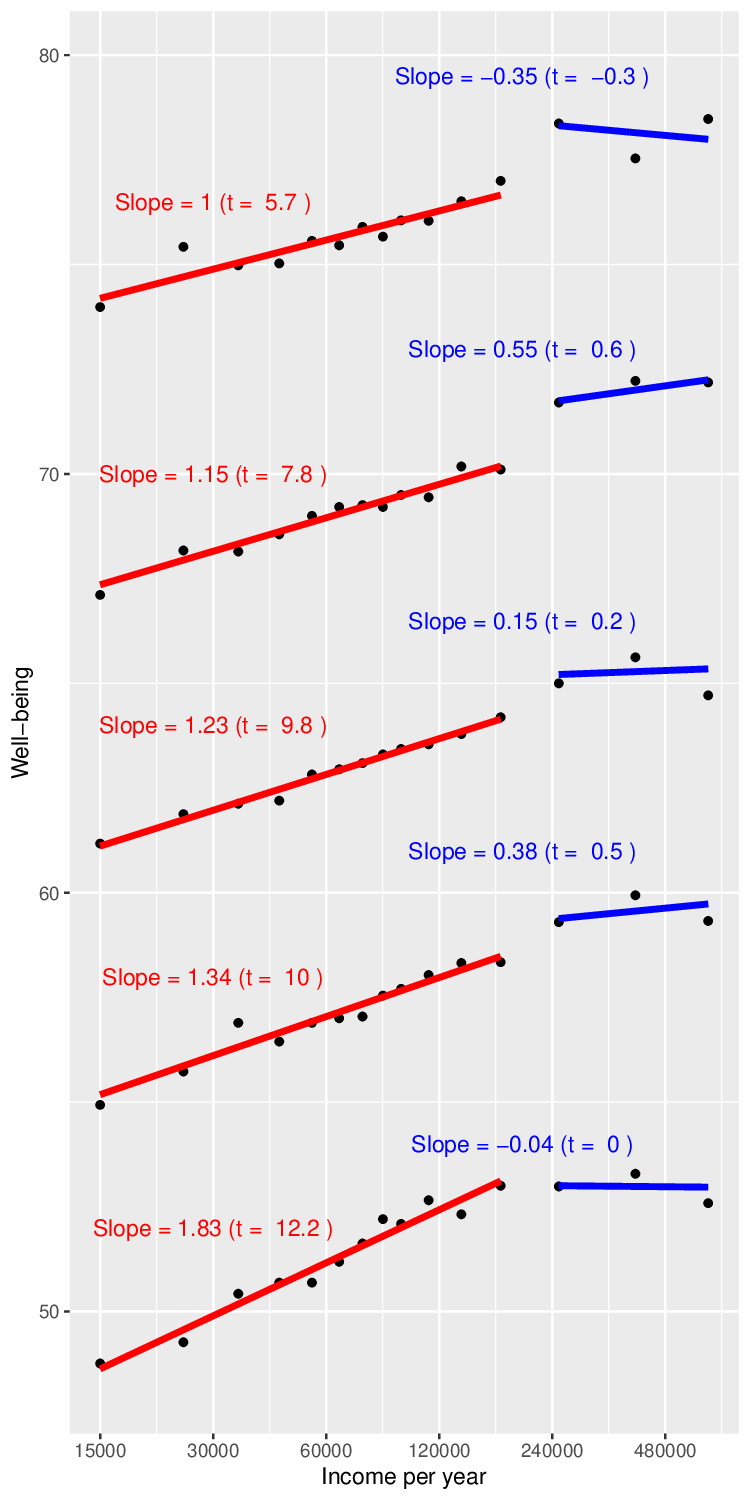}
    \label{fig:quantile}
\end{figure}

For given values of $\tau$ and $p$, the parameters $a,b,c,d$ in \eqref{eq:QR}  may be estimated via standard quantile regression methods \citep[e.g.][]{Koenker2005}. Following \cite{KKM2023}, we consider the quantiles $p = 15\%, 30\%, 50\%, 70\%, 85\%$ of the well-being distribution. The left panel of Figure \ref{fig:quantile} presents the results with the $\tau = $ \$100,000 threshold, which  is a replication of the results from  \cite{KKM2023} and shows a consistently increasing pattern for all but the unhappiest individuals in the sample. For the happiest individuals, the slope even increases above the threshold, pointing to an ``accelerating'' pattern, where the happiest individuals appear to get relatively higher well-being benefits from additional income when their income is already very high. The right panel of Figure \ref{fig:quantile} presents the results with the $\tau = $ \$200,000 threshold suggested by our previous analysis, and shows that choosing the threshold in a data-driven way leads to a flat relationship between income and well-being above the threshold, regardless of the part of the well-being distribution considered. 


\section{Conclusion}\label{sec:concl}

The analysis of this paper challenges  the finding of a monotonically increasing relationship between emotional well-being and income for all but the most unhappy people, recently reported in the literature \citep{KKM2023}. Indeed, our analysis shows that this conclusion rests crucially on pre-specifying the income threshold at \$100,000 per year; by choosing the threshold in a data-driven way, a flat relationship between  emotional well-being and income above a threshold around \$200,000 is found, regardless of the part (i.e. quantile) of the well-being distribution considered. To put this threshold number in perspective, we note that slightly more than 9\% of the sample reported household incomes of \$250,000 or higher. The incomes in the sample matches the US census distribution closely \citep[][]{MK2021}.

The analysis of the present paper, in turn, rests on the assumption that the relationship between emotional well-being and log-income is linear with one structural break in both intercept and slope. While this assumption has been routinely imposed in the literature, it is not given that it is satisfied in practice. Another potential drawback of the analysis conducted here (and elsewhere) is the bracketing of incomes into income groups, which may introduce biases in the results due censoring of the explanatory (income) variable. Further, it is generally believed that well-being depends on a multitude of factors besides income \citep[e.g.][]{Diener2000,Diener2002}, leading to the conjecture that there may be omitted variables in the regression equations \eqref{eq:OLS} and \eqref{eq:QR}. What the effects of these omissions and shortcomings are to the analysis of the relationship between emotional well-being and income remains unknown. Thus, although the analysis of this paper provides some evidence for emotional well-being plateauing around \$200,000 per year, more careful research is needed before any definite conclusions about this relationship can be drawn with confidence.

 {\small 
\bibliographystyle{chicago}
\bibliography{mb_references}
}

\newpage

\clearpage

%
%
%
%

\end{document}